\documentclass[12pt]{article}
\usepackage{amsmath}
\usepackage{amssymb}
\usepackage{latexsym}
\usepackage{graphicx}
\usepackage{epsfig}
\usepackage{pstricks}
\usepackage{cite}

\def\beq{\begin{equation}}
\def\eeq{\end{equation}}
\def\bea{\begin{eqnarray}}
\def\eea{\end{eqnarray}}

\begin{document}

\begin{titlepage}

\vspace*{1cm}
\begin{center}
{\bf \Large On the Localisation of 4-Dimensional\\[2mm] Brane-World Black Holes}

\bigskip \bigskip \medskip

{\bf P. Kanti}, {\bf N. Pappas} and {\bf K. Zuleta}

\bigskip
{\it Division of Theoretical Physics, Department of Physics,\\
University of Ioannina, Ioannina GR-45110, Greece}

\bigskip \medskip
{\bf Abstract}
\end{center}
In the context of brane-world models, we pursue the question of the existence of \mbox{5-dimensional}
solutions describing regular black holes localised close to the brane. Employing a perturbed
Vaidya-type line-element embedded in a warped fifth dimension, we attempt to localise the
extended black-string singularity, and to restore the regularity of the AdS spacetime  
at a finite distance from the brane by introducing an appropriate bulk energy-momentum tensor. 
As  a source for this bulk matter, we are considering a variety of non-ordinary field-theory
models of scalar fields either minimally-coupled to gravity, but including non-canonical kinetic
terms, mixing terms,  derivative interactions and ghosts,  or non-minimally-coupled to
gravity through a general coupling to the Ricci scalar. In all models
considered, even in the ones characterised by a high-degree of flexibility, a negative
result was reached. Our analysis demonstrates how difficult the analytic construction of
a localised brane-world black hole may be in the context of a well-defined field-theory
model. 
Finally, with regard to the question of the existence or not of a static classical
black hole solution on the brane, our analysis suggests that such solutions could in
principle exist, however, the associated field configuration itself has to be dynamic. 
\end{titlepage}


\section{Introduction}
\label{Intro}

The idea that our world could be a 4-dimensional hypersurface, a brane, embedded in a 
higher-dimensional spacetime, the bulk, dates back already to the eighties~\cite{misha, akama}. 
More recently, however, it received a widespread renewed interest when novel theories \cite{ADD, RS} incorporated  gravity into the brane-world scenario in an attempt to solve the hierarchy problem. 
These proposals have prompted an intensive research activity investigating their
implications on gravity, particle physics and cosmology. Gravity, in particular, has seen
one of the most important pillars of the General Theory of Relativity, the concept of
4-dimensional spacetime, being modified in order to accommodate the potential existence
of extra spacelike dimensions. This inevitably led to the reviewing of all known solutions
and predictions of 4-dimensional gravity, the most studied ones being the black-hole
solutions. In the context of the Large Extra Dimensions scenario \cite{ADD},  where the extra
dimensions were assumed to be flat, the study of black holes was straightforward since
higher-dimensional versions of the Schwarzschild \cite{Tangherlini} and Kerr solutions 
\cite{MP} were known for decades. However, in the context of the Warped Extra Dimensions
Scenario \cite{RS}, the task to derive a black hole on a brane embedded in a curved
5-dimensional background has proven to be unexpectedly difficult (for reviews, see
\cite{reviews}).

The first attempt to derive a brane-world black-hole solution in a higher-dimensional
spacetime with a warped extra dimension  appeared in \cite{CHR} where the 
4-dimensional Minkowski line-element in the Randall-Sundrum metric was substituted
by the Schwarzschild solution, i.e
\begin{equation}
ds^2=e^{2A(y)}\,\left[-\left(1-\frac{2M}{r}\right)dt^2 + \left(1-\frac{2M}{r}\right)^{-1} dr^2
+ r^2\,(d\theta^2+\sin^2\theta\,d\varphi^2)\right].
\label{black-string}
\end{equation}
The above line-element satisfies the 5-dimensional Einstein's field equations of the
Randall-Sundrum model since the Schwarzschild solution, just like the Minkowski one, is
a vacuum solution.  However, it was demonstrated that the above ansatz does not
describe a regular black hole localised on the brane since the solution is characterised
by a string-like singularity extended along the fifth dimension. This becomes manifest
in the expression of the 5-dimensional curvature invariant quantity
\bea
R^{MNRS}R_{MNRS} = \frac{48e^{-4A(y)}M^2}{r^6}+...\,.
\label{inv-RS}
\eea
For $A(y)=-k|y|$, where $k$ is the AdS curvature radius, as in the Randall-Sundrum model,
or for any other warp function decreasing away from the brane, the above quantity
blows up at $y$-infinity; more importantly, it reveals the existence of a singularity
at $r=0$ at every slice $y=const$ of the 5-dimensional AdS spacetime. The above
solution was therefore a black string, rather than a black hole, and was soon
proven to be plagued by the Gregory-Laflamme instability \cite{GL, RuthGL}.

In the years that followed, other attempts to derive a regular black-hole solution in a
warped 5-dimensional background proved how tricky the nature of the problem was:
no analytical solution that would satisfy the 5-dimensional field equations and describe
a 4-dimensional black hole on the brane was found, despite numerous different
approaches that were used (for some of them, see \cite{tidal, Papanto, KT, 
KOT, CasadioNew, Frolov, Karasik, GGI, CGKM, Ovalle})\footnote{ In constrast, analytical
solutions describing
black holes localised on a 2-brane embedded in a (3+1)-dimensional bulk were
constructed in \cite{EHM, AS} by using a  C-metric in the AdS$_4$.}.
One of these approaches \cite{KOT} was to assume that the black-hole mass has a non-trivial
$y$-profile along the extra dimension: if $M$ in Eq. (\ref{black-string}) is not a constant
quantity but a function of $y$, then, upon a convenient choice, the
expression on the r.h.s. of Eq. (\ref{inv-RS}) could die out at a finite distance from the
brane. However, the line-element
inside the square brackets in (\ref{black-string}) with $M=M(y)$  is not anymore a vacuum solution. 
A bulk matter distribution must be introduced for the 5-dimensional line-element to
satisfy the field equations. The corresponding energy-momentum tensor was found \cite{KT, KOT}
to describe a shell-like distribution of matter engulfing the brane with a stiff-fluid
equation of state that satisfied all energy conditions on the brane and vanished, as
expected, away from the brane. Unfortunately, no field configuration, in the context of
scalar or gauge field models, was found that could support such an energy-momentum tensor.  

During the same period, numerical solutions were found \cite{KTN, Kudoh, TT} in the context
of five- and six-dimensional warped models that exhibited the existence of black-hole
solutions with horizon radius smaller than or at most of the order of the AdS length
$\ell =1/k$.  No larger black-hole solutions were found, and that led to arguments
of non-existence of large, classical, static black-hole solutions on the brane
\cite{Bruni, Dadhich, Kofinas, Tanaka, EFK, EGK} as well as to counter-arguments
\cite{Fitzpatrick, Zegers, Heydari, Dai}.  
Even in the case of small black holes, no closed-form analytic solutions, that would allow us
to study their topological and physical properties in a comprehensive way, were ever found
-- in addition, the very existence
 of the numerical solutions describing small vacuum black holes was put into
question in recent works \cite{Yoshino, Kleihaus}. Recently, new numerical solutions
employing novel numerical techniques have been presented \cite{FW, Page} that 
describe both small and large black holes in the context of the RS model: the solutions
have been constructed starting from an AdS$_5$/CFT$_4$ solution with an exact
Schwarzschild metric at the AdS infinite boundary; the boundary background is then rewritten in
a more general way and expanded along the bulk to derive a RS brane at a finite proper
distance whose induced metric is a perturbed Schwarzschild metric.     

It is an intriguing fact that, contrary to the findings of the numerical works \cite{KTN, Kudoh, TT},
all analytical attempts to derive a 5-dimensional regular black hole localised on the brane have
been forced to introduce some form of matter in the theory, either in the bulk \cite{KT, KOT, Frolov, Dai}
or on the brane \cite{GGI, CGKM, Ovalle, Heydari, Andrianov}, or even geometrical terms 
\cite{Papanto, Cuadros}. In one of the most recent numerical works that have presented 
brane-localised black-hole solutions \cite{Kleihaus}, the existence of a distribution of matter
also plays an important role: the solutions exist only upon the introduction of
an external electromagnetic field on the brane. In \cite{FW, Page}, as well as in the
lower-dimensional constructions \cite{EHM, AS}, no additional matter is introduced,
however, the induced geometry on the brane is not a vacuum solution of the 4D
equations -- rather, it is sourced by the energy-momentum tensor of the Conformal
Field Theory residing on it (for an introduction to the AdS/CFT correspondence in the
brane-world context, see \cite{HHR}). In our opinion, it is clear that the localisation 
of the black-hole topology -- as we know it -- close to the brane demands support from
some additional form of
matter and cannot be realised by itself. For this reason, in this work, we will turn again to the
approach of \cite{KT, KOT} in order to investigate potential field-theory models that could yield
the well-behaved energy-momentum tensor that supported a regular, localised black hole. 
The mass of the black hole will be assumed again to have a non-trivial profile along the
extra dimension: this is motivated primarily by the need to eliminate the singular term
of Eq. (\ref{inv-RS}) and turn the singular black-string spacetime to a regular AdS one
at a finite distance from the brane; in addition, as the question of whether a purely 
Schwarzschild line-element should be recovered on the brane still remains open, this
$y$-dependence will keep the model general enough to accommodate solutions that
either resemble the Schwarzschild line-element on the brane or deviate from it.
In addition, a time-dependence will be introduced in the line-element in an attempt
to investigate whether the outcome of the gravitational collapse can be indeed static or not.

The outline of our paper is as follows: in section 2, we present the theoretical framework
of our work -- we also make a link with previous analyses \cite{KT, KOT} and justify the
changes in our assumptions. In the following two sections, we proceed to the
investigation of the scalar field-theory models that we have considered in this work: in
section 3, we discuss the first class of models based on one or more scalar fields
minimally-coupled to gravity; in section 4, we turn to the case of a non-minimally-coupled
scalar field with a general coupling to the Ricci scalar. In both classes of models, we investigate the
existence of viable black-hole solutions in the context of a generalised Randall-Sundrum model,
and determine the obstacles that appear while following this approach. 
We discuss our results and present our conclusions in section 5. 

%

\section{The Theoretical Framework}

As mentioned in the Introduction, the factorised metric ansatz (\ref{black-string}) leads to a
black-string{, rather than a black-hole, solution. Therefore, throughout this work, we will
consider a non-factorised metric with a $y$-dependence in the
4-dimensional part of the line-element and more specifically in the mass parameter $M$.
The obvious choice, to substitute the constant $M$ in Eq. (\ref{black-string}) by a function of
the fifth coordinate, however, leads to the appearance of an additional singularity in the
5-dimensional spacetime at the location of the horizon \cite{KT}. In \cite{KOT}, it was
demonstrated that this is due to the non-analyticity of the 4-dimensional line-element:
employing an analytic ansatz, i.e. a 4-dimensional line-element without a horizon, leads
to a 5-dimensional spacetime without additional singularities.

Therefore, in what follows, we will consider the following analytic Vaidya-type line-element
\begin{equation}
\label{metric}
ds^2=e^{2A(y)}\left[-\left(1-\frac{2m(v,y)}{r}\right)dv^2+2\epsilon dv dr+r^2\left(d\theta^2+\sin^2\theta d\varphi^2\right)\right]+dy^2\,.
\end{equation}
For a constant value of $y$, the line-element inside the square brackets is a non-static Vaidya
metric that can be used to describe the dynamical process of a collapsing ($\epsilon=+1$) or
an expanding ($\epsilon=-1$) shell of matter. If we ignore also the $v$-dependence, the
4-dimensional static Vaidya metric is related to the Schwarzschild one by a mere coordinate
transformation. Although we will be interested in final states that describe a static
black hole (thus, we set $\epsilon=+1$), during this work  we will keep the $v$-dependence, 
as we would like to investigate whether static configurations can exist at all or whether
some type of dynamical evolution is necessarily present in the model even after the formation
of the black hole -- as a matter of fact, it was Vaidya-type metrics that were used in some
of the original works addressing this question \cite{Bruni, Dadhich}. 

The modified Vaidya-type line-element (\ref{metric}) was
also shown to exhibit some attractive characteristics in the quest of localised black holes 
\cite{KOT}. Not only is the necessary bulk energy-momentum tensor fairly simple, but 
also the structure of the 5-dimensional spacetime closely resembles that of the factorised
spacetime of the black-string solution -- indeed, the 5-dimensional curvature invariant
quantities for the ansatz (\ref{metric}) have the form
\begin{equation}
R=-20 A'^2 -8 A''\,, \qquad R_{MN} R^{MN}=4 \left(20 A'^4 +16 A'^2 A'' +5 A''^2\right),
\end{equation}

\vskip -4mm
\begin{equation}
R_{MNRS}\,R^{MNRS}=8 \left(5A'^4 +4 A'^2 A'' +2 A''^2 +\frac{6 e^{-4A}\,m^2(v,y)}{r^6} \right),
\label{vaidya_inv}
\end{equation}
and are formally identical to the ones for the metric (\ref{black-string}) with no extra terms
appearing due to the assumed $y$-dependence, a behaviour not observed for any other choice
of non-factorised line-elements. On the other hand, the assumed scaling of the mass function
with $y$ can in principle eliminate the last singular term of 
Eq. (\ref{vaidya_inv}) and restore the
finiteness of the 5-dimensional spacetime at a moderate distance from the brane -- indeed,
any function decreasing faster than the square of the warp factor could achieve 
the localisation
of the black-hole singularity.

The components of the Einstein tensor  $G^M_{\ N}$ for the line-element (\ref{metric})
are found to be:
\bea
&& G^{v}_{\ v} = G^{r}_{\ r}= G^{\theta}_{\ \theta}= G^{\phi}_{\ \phi}=6A'^2 + 3A''\,,
\label{Ein-diag-comp}\\[1mm]
&&G^{r}_{\ v} = \frac{2}{r^2}\,e^{-2A}\,\partial_{v}m  - \frac{1}{r}\,(\partial^2_{y}m +
4A'\,\partial_{y}m)\,, \label{Ein-rv-comp} \\[1mm]
&& G^{y}_{\ v} = e^{2A}\,G^{r}_{\ y}= \frac{1}{r^2}\,\partial_{y}m\,, \label{Ein-yv-comp} \\[1.5mm]
&& G^{v}_{\ r}=G^{y}_{\ r} =G^{v}_{\ y} =0\,, \label{Ein-van-comp} \\[1.5mm]
&& G^{y}_{\ y} = 6A'^2\,. \label{Ein-yy-comp}
\eea
The Einstein's field equations in the bulk will follow by equating the above components
of $G^M_{\ N}$ with the corresponding ones of the energy-momentum tensor $T^M_{\ N}$. 
The latter will be determined once the bulk Lagrangian is defined, in the next section. 
However, the form of the above Einstein tensor components allows us to make some basic observations.
The assumed $y$-dependence of the mass function introduces off-diagonal, non-isotropic
pressure components. The dependence on $v$ does not by itself introduce a new pressure
component but contributes to one of the non-isotropic ones. In \cite{KOT}, the assumption
was made that the warp factor has the form of the Randall-Sundrum model, $A(y)=-k|y|$,
which is supported by the bulk cosmological constant. In that case, the Einstein equations 
corresponding to the diagonal components (\ref{Ein-diag-comp}) and (\ref{Ein-yy-comp}) are 
trivially satisfied and no energy density
or diagonal pressure components are necessary in the bulk. Here, however, we will assume
that the warp factor has a general form $A(y)$ in order to allow for less restricted field
configurations that, in general, generate both diagonal and off-diagonal components.
Since~$G^{v}_{\ v} = G^{r}_{\ r}= G^{\theta}_{\ \theta}= G^{\phi}_{\ \phi}$,
the bulk energy-momentum tensor will satisfy, by construction, a stiff equation of state.

In the following sections, we will study a variety of field theory models in an attempt to
find the one that could support the aforementioned line-element (\ref{metric}). It is already
known \cite{KT} that the desired Vaidya-type metric cannot be supported by conventional forms
of matter (realised by either scalar or gauge fields). Motivated by previous considerations
of non-ordinary scalar field theories, that aimed to produce additional pressure 
components necessary for the stabilisation of brane-world models \cite{KKOP, Csaki, KOP}, we will
focus our attention on scalar fields and consider a variety of models. These will include
one or more scalar fields minimally coupled to gravity but with a general Lagrangian, admitting
non-canonical kinetic terms, derivative interactions, mixing terms or the presence of 
ghosts, as well as a scalar field non-minimally-coupled to gravity with a general
coupling to the Ricci scalar.

Once a consistent solution in the bulk is found, a single brane will then be introduced
in the model that in general contains a localised energy-momentum tensor
$S_{\mu \nu}$. The spacetime will be assumed to be invariant under the mirror
transformation $y \rightarrow -y$. The bulk equations will then be supplemented by
the junction conditions \cite{Israel} 
\begin{equation}
[K_{\mu\nu} - h_{\mu\nu}\,K]=-\kappa^2_5\,S_{\mu\nu}\,,
\end{equation}
relating the extrinsic curvature $K_{\mu\nu}$, the induced metric tensor $h_{\mu\nu}$
and the energy-momentum tensor $S_{\mu \nu}$ on the brane - the brackets denote
the discontinuity across the brane. The discontinuity of the l.h.s. of the above equation
will be a function of the warp factor $A(y)$, the mass function $m(v,y)$ and their
derivatives with respect to $y$. With the help of the bulk solution, if existent, the
above equation will give us the necessary matter content of the brane for its consistent
embedding in the 5-dimensional warped spacetime.   

%

\section{A Field Theory with minimally-coupled Scalars} \label{minimal-gen}

In this section, we focus on the case of models with minimally-coupled scalar
fields with a general form of Lagrangian. The action functional of the gravitational
theory therefore reads
\bea
\mathcal{S} = \int d^4x\,dy\,\sqrt{-g} \,\left(\frac{R}{2 \kappa_5^2}-
\mathcal{L}_{sc}-\mathcal{L}_{m}\right), \label{action-min}
\eea
where $g_{MN}$ and $R$ are the metric tensor and Ricci scalar, respectively, of the
5-dimensional spacetime described by (\ref{metric}), and $\kappa_5^2=8\pi G_N$ the
5-dimensional gravitational constant.
The action contains in addition the general Lagrangian $\mathcal{L}_{sc}$, associated
with one or more scalar fields, and $\mathcal{L}_{m}$ stands for any other 
form of matter or energy in the theory - throughout this work, we will assume that
this term describes the distribution of a uniform, negative energy-density and
thus $\mathcal{L}_{m}=\Lambda_B$, where $\Lambda_B$ the bulk cosmological constant.
The field equations resulting from the aforementioned action have the form
\bea
R_{MN}-\frac{1}{2}\,g_{MN}\,R = 
\kappa^2_{5}\,(T_{MN} -g_{MN}\,\Lambda_B)\,, \label{eq-motion}
\eea
with $T_{MN}$ being the energy-momentum tensor associated with the scalar fields
\beq
T_{MN}=\frac{2}{\sqrt{-g}}\,\frac{\delta (\sqrt{-g}\,\mathcal{L}_{sc})}
{\delta g^{MN}}\,.
\eeq
In the following subsections, we consider particular choices for $\mathcal{L}_{sc}$,
and we examine the existence of a viable solution of the field equations in the bulk.


\subsection{A Single Scalar Field with a non-canonical kinetic term} \label{single}

As a fist step, we consider the following theory of a single scalar field with a
non-canonical kinetic term
\begin{equation}
\mathcal{L}_{sc}=\sum_{n=1}\,f_n(\phi)\left(\partial^M\phi\,\partial_M\phi\right)^n+V(\phi)\,,
\end{equation}
where $f_n(\phi)$ are arbitrary, smooth functions of the scalar field $\phi$.
The components of the corresponding energy-momentum tensor follow from the expression
\begin{equation}
T^A_{\ B}=2\sum_{n=1}nf_n(\phi)\left(\partial^M\phi\,\partial_M\phi\right)^{n-1}\partial^A\phi\,\partial_B\phi-
\delta^A_B  \,\mathcal{L}_{sc}.
\end{equation}
The off-diagonal components $T^v_{\ r}$, $T^y_{\ r}$ and $T^v_{\ y}$  of the energy-momentum
tensor must trivially vanish since the corresponding components of the Einstein tensor, Eq.
(\ref{Ein-van-comp}), do the same. These conditions however impose strict constraints on the form
of the scalar field: the vanishing of the $T^v_{\ r}$ component, for instance,
\begin{eqnarray}
T^v_{\ r}=
2\sum_{n=1}nf_n(\phi)\left(\partial^M\phi\,\partial_M\phi\right)^{n-1}(\partial_r\phi)^2\,e^{-2A} 
\end{eqnarray}
demands that the scalar field be not a function of the radial coordinate, $\partial_r\phi=0$.
But then it is not possible to satisfy the remaining Einstein's equations: assuming that 
$\phi=\phi(v,y)$\footnote{Throughout this work, and in order to preserve the spherical symmetry
of any potential solution, we assume that the scalar fields do not depend on the angular
coordinates $\theta$ and $\phi$.}, the expression of the non-vanishing off-diagonal component
$T^y_{\ v}$, when combined with the corresponding component of the Einstein tensor
(\ref{Ein-yv-comp}), leads to the equation
\begin{equation}
\frac{\partial_y m}{r^2} = 2\kappa^2_5\sum_n n f_n(\phi)\,(\partial_y\phi)^{2n-1}\,\partial_v\phi\,.
\end{equation}
An incompatibility problem arises immediately: the field $\phi$ and, therefore, the
right-hand-side of the above equation is independent of $r$ but the left-hand-side
has an explicit dependence on that coordinate. As a result, the case of a single, minimally-coupled 
scalar field, even with a general non-canonical kinetic term, does not lead to a solution.


\subsection{Two interacting scalar fields}

We are thus forced to consider a multi-field model. We will study first the case of
two scalar fields $\phi$ and $\chi$ whose dynamics and interactions are described by
the Lagrangian
\begin{equation}
\label{action_two_1}
\mathcal{L}_{sc}=f^{(1)}(\phi,\chi)\,\partial^M\phi\,\partial_M\phi+
f^{(2)}(\phi,\chi)\,\partial^M\chi\,\partial_M\chi+V(\phi,\chi)\,,
\end{equation}
where $f^{(1,2)}$ are arbitrary smooth functions of the two fields. The energy-momentum tensor
now reads:
\begin{equation}
T^A_{\ B}=2f^{(1)}(\phi,\chi)\,\partial^A\phi\,\partial_B\phi+
2f^{(2)}(\phi,\chi)\,\partial^A\chi\,\partial_B\chi-\delta^A_{\ B}\,\mathcal{L}_{sc}\,.
\label{Tmn-2scalars}
\end{equation}

The vanishing of the off-diagonal components $G^v_{\ r}$, $G^y_{\ r}$ and $G^v_{\ y}$ 
implies again the vanishing of the corresponding components of the energy-momentum tensor,
which now results in the following two constraints\footnote{$T^v_{\ y}=g^{vr}\,T^y_{\ r}$,
and as a result there are only two independent constraints.} on the fields:
\begin{eqnarray}
\label{constr_1}
&&f^{(1)}(\phi,\chi)\,(\partial_r\phi)^2+f^{(2)}(\phi,\chi)\,(\partial_r\chi)^2=0\,, \\[1mm]
\label{constr_2}
&&f^{(1)}(\phi,\chi)\,\partial_r\phi\partial_y\phi+f^{(2)}(\phi,\chi)\,\partial_r\chi\partial_y\chi=0\,. 
\end{eqnarray}             
From the constraint (\ref{constr_1}), it is clear that if one of the fields were not to depend
on~$r$, neither would the other one. Although in this case both constraints would be
trivially satisfied, the $(yv)$-component of the field equations, which now has the form
\begin{equation}
\frac{\partial_y m}{r^2}=2\kappa_5^2 \left[f_1^{(1)}(\phi,\chi)\partial_y\phi\partial_v\phi+f_1^{(2)}(\phi,\chi)\partial_y\chi\partial_v\chi\right],
\label{G_yv}
\end{equation}  
would again present an inconsistency, the r.h.s. being necessarily $r$-independent and the
l.h.s. a function of $r$.
Similarly, the constraint (\ref{constr_2}) implies that if one of the fields were not to depend on $y$,
neither would the other one. But this case is also excluded since, through Eq.~(\ref{G_yv}), the mass
of the black hole would then necessarily loose the assumed $y$-dependence. 

The constraints (\ref{constr_1})-(\ref{constr_2}) are supplemented by a third one following from
the diagonal components of the Einstein's field equations along the brane. By using the
expression (\ref{Tmn-2scalars}) and applying the constraint (\ref{constr_1}), the corresponding
components of the energy-momentum tensor are found to have the form:
\begin{eqnarray}
&~& T^v_{\ v}=T^r_{\ r}=2 e^{-2A}\,\Bigl[f^{(1)}(\phi,\chi)\,\partial_r\phi \partial_v\phi+
f^{(2)}(\phi,\chi)\,\partial_r\chi \partial_v\chi\Bigr] -\mathcal{L}_{sc}\,,  \label{t_vv}\\[1mm]
&~&  T^\theta_{\ \theta}=T^\varphi_{\ \varphi}=-\mathcal{L}_{sc}\,. \label{t_pp}
\end{eqnarray}
The components of the Einstein tensor along the brane, given in Eq. (\ref{Ein-diag-comp}),
satisfy the relation $G^{v}_{\ v} = G^{r}_{\ r}= G^{\theta}_{\ \theta}= G^{\phi}_{\ \phi}$,
therefore the corresponding components of $T^M_{\ N}$ should also be equal. Comparing (\ref{t_vv}) and (\ref{t_pp}), it is obvious that this holds if an additional constraint on the field configurations is imposed, namely
\begin{equation}
f^{(1)}(\phi,\chi)\,\partial_r\phi \partial_v\phi+
f^{(2)}(\phi,\chi)\,\partial_r\chi \partial_v\chi=0\,. \label{constr_3}
\end{equation}
From the above constraint, we may again conclude that if one of the fields were not to depend
on $v$, neither would the other one. However, we note from Eq. (\ref{G_yv}), that, in
order for a solution with a non-trivial profile of the mass distribution $m=m(y)$ to exist,
the fields must necessarily depend on $v$. In other words, if such a solution exists,
the matter distribution around a black hole must be dynamical and not static, 
even if the mass of the black hole itself is not time-evolving and thus independent of $v$.

Coming back to the existence of the solution and assuming that $\phi=\phi(v,r,y)$ and 
$\chi=\chi(v,r,y)$, we proceed as follows: we solve the new constraint (\ref{constr_3}) for
the coupling function $f^{(2)}(\phi,\chi)$, and then substitute it into the $(yv)$-component
(\ref{G_yv}) to obtain the following alternative form for that equation
 \begin{equation}
\frac{\partial_y m}{r^2}=2\kappa_5^2\,f^{(1)}(\phi,\chi)\,\frac{\partial_v \phi}{\partial_r \chi}
\left(\partial_y\phi\partial_r\chi-\partial_y\chi\partial_r\phi\right).
\label{G_yv-new}
\end{equation}
However, a similar rearrangement of Eq. (\ref{constr_1}) and substitution into the constraint
(\ref{constr_2}) leads to 
\begin{equation}
\partial_y\phi\partial_r\chi-\partial_y\chi\partial_r\phi=0\,,
\end{equation}
that unfortunately causes the r.h.s. of Eq. (\ref{G_yv-new}) to be zero and thus the mass function looses the
desired $y$-dependence. We note that the absence of the solution holds
independently of the signs of the coupling functions $f^{(1,2)}(\phi,\chi)$ --
i.e. of whether the two scalars are positive-norm fields or whether they are ghosts --
or of the form of the potential $V(\phi,\chi)$ that determines the interaction between the two fields.


\subsection{Two interacting scalar fields  with general kinetic terms}

We now combine the two previous models considered to construct a Lagrangian of two scalar
fields interacting through an arbitrary potential $V(\phi,\chi)$ and admitting general non-canonical
kinetic terms. The Lagrangian of the scalar fields then reads
\begin{equation}
{\cal L}_{sc}= \sum_{n=1}f_n^{(1)}(\phi,\chi)\left(\partial^M\phi\partial_M\phi\right)^n+
\sum_{n=1}f_n^{(2)}(\phi,\chi)\left(\partial^M\chi\partial_M\chi\right)^n+V(\phi,\chi)\,,
\end{equation}
while the energy momentum tensor assumes the form:
\begin{eqnarray}
T^A_{\ B}&=&2\sum_{n=1}f^{(1)}_n(\phi,\chi)\,n\left(\partial^M\phi\partial_M\phi\right)^{n-1}
\partial^A\phi\partial_B\phi \nonumber \\[1mm]
&& \hspace*{2cm} + 2\sum_{n=1}f^{(2)}_n(\phi,\chi)\,n\left(\partial^M\chi\partial_M\chi\right)^{n-1}\partial^A\chi\partial_B\chi 
-\delta^A_{\ B}\,{\cal L}_{sc}\,. \label{Tmn-2non}
\end{eqnarray}

Working as in the previous subsection, from the vanishing of the off-diagonal components
$G^v_{\ r},\, G^v_{\ y}$ and~$G^y_{\ r}$, we derive the following two constraints on the fields 
\begin{eqnarray}
\label{constr_1_v2}
&& \hspace*{-1cm}
\sum_{n=1}n\left[f_n^{(1)}(\phi,\chi)\left(\partial^M\phi\partial_M\phi\right)^{n-1}(\partial_r\phi)^2
+f_n^{(2)}(\phi,\chi)\left(\partial^M\chi\partial_M\chi\right)^{n-1}(\partial_r\chi)^2\right]=0\,, \\[1mm]
\label{constr_2_v2}
&& \hspace*{-1cm}
\sum_{n=1}n\left[f_n^{(1)}(\phi,\chi)\left(\partial^M\phi\partial_M\phi\right)^{n-1}\partial_r\phi\partial_y\phi
+f_n^{(2)}(\phi,\chi)\left(\partial^M\chi\partial_M\chi\right)^{n-1}\partial_r\chi\partial_y\chi\right]=0\,.
\end{eqnarray} 
Also, the equality of the diagonal components of the Einstein tensor along the brane results
into the additional constraint
\begin{equation}
\sum_{n=1}n\left[f_n^{(1)}(\phi,\chi)\left(\partial^M\phi\partial_M\phi\right)^{n-1}\partial_r\phi \partial_v\phi+
f_n^{(2)}(\phi,\chi)\left(\partial^M\chi\partial_M\chi\right)^{n-1}\partial_r\chi \partial_v\chi\right]=0\,, \label{constr_3_v2}
\end{equation}
while the $(yv)$-component of the Einstein's field equations now has the form
\begin{equation}
\label{G_yv_v2}
\frac{\partial_y m}{r^2}=2\kappa_5^2\sum_{n=1}n \left[f_n^{(1)}(\phi,\chi)
\left(\partial^M\phi\partial_M\phi\right)^{n-1} \partial_y\phi\partial_v\phi+
f_n^{(2)}(\phi,\chi)\left(\partial^M\chi\partial_M\chi\right)^{n-1}\partial_y\chi\partial_v\chi\right].
\end{equation}  
The following observation makes the attempt to find a viable solution in the context of this
model obsolete: if we define the following functions
\begin{eqnarray}
\tilde f^{(1)}(\phi,\chi) &=&
\sum_{n=1}n f_n^{(1)}(\phi,\chi) \left(\partial^M\phi\partial_M\phi\right)^{n-1}\,, \\
\tilde f^{(2)}(\phi,\chi)&=&
\sum_{n=1}n f_n^{(2)}(\phi,\chi) \left(\partial^M\chi\partial_M\chi\right)^{n-1}\,,
\end{eqnarray}
then, Eqs. (\ref{constr_1_v2}-\ref{G_yv_v2}) reduce to Eqs. (\ref{constr_1}), (\ref{constr_2}), (\ref{constr_3}),
and (\ref{G_yv}), respectively, with the $f^{(1,2)}(\phi,\chi)$ coupling functions being
replaced by $\tilde f^{(1,2)}(\phi,\chi)$. As a result, upon a similar rearrangement of the
three constraints, the r.h.s. of the $(yv)$-component vanishes, a result that eliminates again the
$y$-dependence of the mass function. 
            

\subsection{Two interacting scalar fields with mixed kinetic terms\label{section_mixing_1}}

We now increase the complexity of the model by allowing the scalar fields to have mixed
kinetic terms and thus consider the following generalized form of the scalar Lagrangian
\begin{equation}
\label{action_two_1_mixed}
{\cal L}_{sc}=f^{(1)}(\phi,\chi)\,\partial^M\phi\partial_M\phi+
f^{(2)}(\phi,\chi)\,\partial^M\chi\partial_M\chi+f^{(3)}(\phi,\chi)\,\partial^M\phi\partial_M\chi
+V(\phi,\chi)\,.
\end{equation}
Then, the energy-momentum tensor reads:
\begin{eqnarray} \hspace*{-0.5cm}
T^A_{\ B}&=&2f^{(1)}(\phi,\chi)\,\partial^A\phi\partial_B\phi+2f^{(2)}(\phi,\chi)\,\partial^A\chi\partial_B\chi
\nonumber \\[1mm]
&& \hspace*{1cm} +\,f^{(3)}(\phi,\chi)\left[\partial^A\phi\partial_B\chi+\partial^A\chi\partial_B\phi\right]
-\delta^A_{\ B}\,{\cal L}_{sc}\,.
\end{eqnarray}
The vanishing of the off-diagonal components $G^v_{\ r},\, G^v_{\ y}$ and~$G^y_{\ r}$ imposes again
the vanishing of the corresponding components of the energy-momentum tensor, which in this case
results in the following two constraints
\begin{eqnarray}
\label{constr_1_mixed}
&& \hspace*{-1cm}
f^{(1)}(\phi,\chi)\,(\partial_r\phi)^2+f^{(2)}(\phi,\chi)\,(\partial_r\chi)^2+f^{(3)}(\phi,\chi)\,\partial_r\phi\partial_r\chi=0\,, \\[1mm]
\label{constr_2_mixed}
&&  \hspace*{-1cm}
2f^{(1)}(\phi,\chi)\,\partial_r\phi\partial_y\phi+2f^{(2)}(\phi,\chi)\,\partial_r\chi\partial_y\chi+
f^{(3)}(\phi,\chi)\left[\partial_r\phi\partial_y\chi+\partial_y\phi\partial_r\chi\right]=0\,. 
\end{eqnarray}        
From the first of the above two equations, it is clear that either both fields must simultaneously
depend on the radial coordinate $r$ or they must both be $r$-independent. If they are both
independent of $r$, then the two constraints are satisfied, but the non-vanishing off-diagonal
$(yv$)-component, that now takes the form
\begin{equation}
\frac{\partial_y m}{r^2}= \kappa^2_5\left[2f^{(1)}\,\partial_v\phi\partial_y\phi+
2f^{(2)}\,\partial_v\chi\partial_y\chi+
f^{(3)}\left(\partial_v\phi\partial_y\chi+\partial_y\phi\partial_v\chi\right)\right],
\label{G_yv_mixed}
\end{equation}
becomes inconsistent due to the explicit $r$-dependence on its l.h.s.. Equation 
(\ref{G_yv_mixed}) seems to allow for certain combinations of the partial derivatives
 ($\partial_y \phi$, $\partial_y \chi$, $\partial_v \phi$, $\partial_v \chi$) to vanish. 
However, in what follows, we will assume the most general case, i.e. that $\phi=\phi(r,v,y)$
and $\chi=\chi(r,v,y)$, and we will comment on particular cases at the end of this subsection.

We now turn to the diagonal components of the Einstein's field equations. The diagonal components
of the Einstein tensor along the brane are equal, and thus the same must hold for the components of the
energy-momentum tensor, that now have the form
\begin{eqnarray}
&& \hspace*{-1cm}
T^v_{\ v}=T^r_{\ r}=e^{-2A} \left[2f^{(1)}\partial_r\phi\partial_v\phi+2f^{(2)}\partial_r\chi\partial_v\chi+
f^{(3)}\left(\partial_r\phi\partial_v\chi+\partial_r\chi\partial_v\phi\right)\right]-{\cal L}_{sc}\,,\\[1mm]
&& \hspace*{-1cm}T^\theta_{\ \theta}=T^\varphi_{\ \varphi}=-{\cal L}_{sc}\,. 
\end{eqnarray}
Demanding the equality of the above expressions, the following additional constraint is obtained
\begin{equation}
2f^{(1)}(\phi,\chi)\,\partial_r\phi\partial_v\phi+2f^{(2)}(\phi,\chi)\,\partial_r\chi\partial_v\chi+
f^{(3)}(\phi,\chi)\left(\partial_r\phi\partial_v\chi+\partial_r\chi\partial_v\phi\right)=0\,. \label{constr_3_mixed}
\end{equation}

Let us now consider the system of constraints (\ref{constr_1_mixed}), (\ref{constr_2_mixed}) and 
(\ref{constr_3_mixed}): it is a homogeneous system of linear equations for $f^{(1)}$, $f^{(2)}$ and
$f^{(3)}$ -- the necessary condition for this system to possess a solution other than the trivial
one is the vanishing of the determinant of the matrix of coefficients:

\begin{equation}
\left|
\begin{array}{clcr}
(\partial_r\phi)^2 & (\partial_r\chi)^2 & \partial_r\phi\partial_r\chi\\[1mm]
2\partial_r\phi\partial_y\phi & 2\partial_r\chi\partial_y\chi &\partial_r\phi\partial_y\chi+
\partial_y\phi\partial_r\chi \\[1mm]
2\partial_r\phi\partial_v\phi & 2\partial_r\chi\partial_v\chi &\partial_r\phi\partial_v\chi+
\partial_v\phi\partial_r\chi\end{array}\right|=0\,.
\end{equation}
One may easily check that the above condition indeed holds and therefore the system may be solved
to yield the values of two coupling functions in terms of the third one. In this way, we find:
\begin{equation}
f^{(1)}=f^{(2)}\,\frac{(\partial_r \chi)^2}{(\partial_r \phi)^2}\,, \qquad \qquad 
f^{(3)}=-2 f^{(2)}\,\frac{\partial_r \chi}{\partial_r \phi}\,.
\label{coupl_mixed}
\end{equation}
If we then use the above relations in the expression of the $(yv)$-component (\ref{G_yv_mixed}),
we obtain the alternative form
\begin{equation}
\frac{\partial_y m}{r^2}= \frac{2\kappa^2_5 f^{(2)}}{(\partial_r \phi)^2}
\left(\partial_v\phi\partial_r\chi-\partial_r\phi\partial_v\chi\right)
\left(\partial_y\phi\partial_r\chi-\partial_r\phi\partial_y\chi\right).
\label{G_yv_mixed_alter}
\end{equation}
We observe that, contrary to what happens in the previous two models considered, the rearrangement
of the three constraints (\ref{constr_1_mixed}), (\ref{constr_2_mixed}) and (\ref{constr_3_mixed}) in
this model does not by itself cause the vanishing of the r.h.s. of the above equation. Clearly,
as the Lagrangian of the model becomes more complex, the system of field equations becomes
more flexible.  

The remaining independent off-diagonal component that we have not considered yet follows by
combining the $G^r_{\ v}$ component (\ref{Ein-rv-comp}) of the Einstein tensor with the
corresponding component of the energy-momentum tensor. Then, we obtain the field equation
\begin{equation}
\frac{2\partial_v m}{r^2}-\frac{e^{2A}}{r} \left(\partial^2_y m+ 4A' \partial_y m\right)=
2\kappa^2_5\left[f^{(1)}\,(\partial_v\phi)^2+f^{(2)}\,(\partial_v\chi)^2+
f^{(3)}\,\partial_v\phi\partial_v\chi\right].
\label{G_rv_mixed}
\end{equation}
Similarly, if we use the relations (\ref{coupl_mixed}) in the above equation, this may be
rewritten as
\begin{equation}
\frac{2\partial_v m}{r^2}-\frac{e^{2A}}{r} \left(\partial^2_y m+ 4A' \partial_y m\right)=
2\kappa^2_5\frac{f^{(2)}}{(\partial_r\phi)^2}\,(\partial_v \phi \partial_r \chi-
\partial_r \phi \partial_v \chi)^2\,.
\label{G_rv_mixed_v2}
\end{equation}

Finally, the last diagonal component, the ($yy$)-component, assumes the form
\begin{equation}
6 A'^2 = \kappa_5^2 \left[-\Lambda_B + 2 f^{(1)} (\partial_y \phi)^2
+ 2 f^{(2)} (\partial_y \chi)^2 + 2 f^{(3)} \partial_y \phi \partial_y\chi -{\cal L}_{sc}\right].
\label{yy-eq}
\end{equation}
At this point we will need the explicit expression of the Lagrangian ${\cal L}_{sc}$. 
By making use of the constraints (\ref{constr_1_mixed}) and (\ref{constr_3_mixed}),
this turns out to be 
\begin{equation}
{\cal L}_{sc} = f^{(1)} (\partial_y \phi)^2 + f^{(2)} (\partial_y \chi)^2 +
f^{(3)} \,\partial_y \phi \partial_y \chi + V(\phi, \chi)\,.
\label{Lagr-mixed}
\end{equation}
If we use the above expression, then Eq. (\ref{yy-eq}) and the diagonal components of the
field equations along the brane reduce to the following two independent differential equations
\begin{eqnarray}
6 A'^2 &=& \kappa_5^2 \left[-\Lambda_B + f^{(1)} (\partial_y \phi)^2
+ f^{(2)} (\partial_y \chi)^2 + f^{(3)} \partial_y \phi \partial_y\chi -V(\phi, \chi)\right],
\label{yy-eq-new} \\[1mm]
6 A'^2 + 3A'' &=& \kappa_5^2 \left[-\Lambda_B - f^{(1)} (\partial_y \phi)^2
- f^{(2)} (\partial_y \chi)^2 - f^{(3)} \partial_y \phi \partial_y\chi -V(\phi, \chi)\right],
\label{mm-eq}
\end{eqnarray}
respectively. Subtracting the first of the above equations from the second, the
latter takes the  simpler form
\begin{equation}
3A'' = -2 \kappa_5^2 \left[f^{(1)} (\partial_y \phi)^2
+ f^{(2)} (\partial_y \chi)^2 + f^{(3)} \partial_y \phi \partial_y\chi\right]= 
-\frac{ 2  \kappa_5^2\,f^{(2)}}{(\partial_r\phi)^2}\,(\partial_y \phi \partial_r \chi-
\partial_r \phi \partial_y \chi)^2\,,
\label{mm-eq-new}
\end{equation}
where, in the last part, we have used again the relations (\ref{coupl_mixed}). If we now take the
square of Eq. (\ref{G_yv_mixed_alter}) and combine it with Eqs. (\ref{G_rv_mixed_v2}) and
(\ref{mm-eq-new}), we obtain a differential equation for the mass function with no dependence
on the fields and their coupling functions, namely
\begin{equation}
\frac{(\partial_y m)^2}{r^3}= 3A'' \left[
-\frac{2\partial_v m}{r}+e^{2A}\left(\partial^2_y m+ 4A' \partial_y m\right)\right]\,.
\end{equation}
Unfortunately, this equation is again inconsistent as it involves an explicit dependence on the
radial coordinate on which the mass function is assumed not to depend.

In addition to the above, this model is plagued by another problem following from
the restrictions that the field equations
impose on the field configurations: as the warp factor is solely a function of the $y$-coordinate,
then through Eqs. (\ref{mm-eq-new}) and (\ref{yy-eq-new}), the potential $V(\phi, \chi)$ should
also be a function of $y$. Assuming that the
potential depends on both fields, and that these are general functions of the $(r,v,y)$
coordinates, $V(\phi, \chi)$ ought to have a particular form so that its dependence
on the $(r,v)$ coordinates vanishes. These forms could be:

\begin{itemize}
\item[$\bullet$] $V(\phi, \chi)=F(\chi^n+\phi^n)$, where $n$ an arbitrary integer and $F$ an
arbitrary function of the combination $\chi^n+\phi^n$. For the latter to be a function of $y$,
we should also have:
$\chi^n=\chi_1(y) +\chi_2(r,v)$ and $\phi^n=\phi_1(y) + \phi_2(r,v)$, with 
$\phi_2(r,v) =-\chi_2(r,v)$. But then, one may easily show that
\begin{equation}
\partial_r\chi \partial_v\phi-\partial_r\phi \partial_v\chi= \frac{(\chi \phi)^{1-n}}{n^2}\,
\left(\partial_v \chi_2 \partial_r\chi_2-\partial_r \chi_2 \partial_v\chi_2\right)=0\,,
\label{pot-1}
\end{equation}
in which case the r.h.s. of the $(yv)$-component (\ref{G_yv_mixed_alter}), and the assumed
dependence of the mass function on $y$, vanishes.

\item[$\bullet$] $V(\phi, \chi)=G(\chi^{n_1}\phi^{n_2})$, where $G$ an arbitrary function of the
combination $\chi^{n_1}\phi^{n_2}$, and $(n_1, n_2)$ arbitrary integers. In this case, we should
have: $\chi=\chi_1(y)\chi_2(r,v)$ and $\phi=\phi_1(y)\phi_2(r,v)$, with 
$\phi_2(r,v) =c \chi_2(r,v)^{-n_1/n_2}$ and $c$ a constant. Once again, the
combination $(\partial_r\chi \partial_v\phi-\partial_r\phi \partial_v\chi)$ is easily 
found to be zero. 
\end{itemize} 

Let us finally investigate whether more specific assumptions on the form of the fields
or the potential are allowed. Clearly, the case where the potential $V$ depends only on
one of the two fields, i.e. $V=V(\chi)$, is excluded:  $\chi$ must necessarily depend
on $r$, as discussed below Eqs. (\ref{constr_1_mixed})-(\ref{constr_2_mixed}), and
the presence of $\phi$ in the
expression of the potential is imperative in order for this $r$-dependence to cancel.
The same argument excludes the case where only one of the two fields depend on the 
time-coordinate $v$, as in that case $V(\chi,\phi$) would carry this $v$-dependence.
The case where none of the two fields depend on $v$ is also rejected since then 
the r.h.s. of Eq. (\ref{G_yv_mixed_alter}) would be zero - the same holds if we
assume that both fields are not functions of the extra coordinate. The assumption
that only one of the two fields may depend on $y$ is the only one allowed with
Eqs. (\ref{G_yv_mixed_alter}), (\ref{yy-eq-new}) and (\ref{mm-eq-new}) assuming
then simpler, yet non-trivial forms -- nevertheless, this assumption does not 
alter the arguments presented above regarding the form of the potential and thus
fails to lead to a viable solution. 

The analysis presented in this subsection may be easily generalised to allow for more
general kinetic terms along the lines of subsection 3.3. Then, the Lagrangian would read
\begin{eqnarray}
&& \hspace*{-2.0cm}{\cal L}_{sc}=\sum_{n=1}\left[f_n^{(1)}(\phi,\chi)\left(\partial^M\phi\partial_M\phi\right)^n+
f_n^{(2)}(\phi,\chi)\left(\partial^M\chi\partial_M\chi\right)^n \right. \nonumber \\[1mm]
&&\hspace*{3.5cm}\left.+f_n^{(3)}(\phi,\chi)\left(\partial^M\phi\partial_M\chi\right)^n\right]+V(\phi,\chi)\,.
\end{eqnarray}
Although the expressions of all constraints and non-vanishing field equations would
become more complicated, one may again show that these, upon conveniently redefining the
coupling functions, reduce to the ones presented in this subsection.
As the same arguments regarding the restrictions on the potential and form of fields
would still hold, no viable solution would emerge in the context of this model either.


\section{A non-minimally-coupled Scalar Field Theory}
\label{Conformal}

Let us now turn to the case of a scalar-tensor theory of gravity with a non-minimally coupled
scalar field present in the bulk. We consider the following general form of the action
\bea
\mathcal{S} = \int d^4x\,dy\,\sqrt{-g} \,\left[\frac{f(\Phi)}{2 \kappa_5^2}\,R-
\frac{1}{2}(\nabla\Phi)^2-V(\Phi)-\Lambda_B\right]\,, \label{action}
\eea
where $f(\Phi)$ is an arbitrary, smooth, positive-definite function of the scalar
field $\Phi$, and $g_{MN}$ is the five-dimensional metric given again by Eq. (\ref{metric}).
The equations of motion resulting from the aforementioned action have the form
\bea
f(\Phi)\bigl(R_{MN}-\frac{1}{2}\,g_{MN}\,R\bigr) = 
\kappa^2_{5}\,(-g_{MN}\,\Lambda_B + \mathbb{T}^{(\Phi)}_{MN})\,, \label{eq-motion-conf}
\eea
with $\mathbb{T}^{(\Phi)}_{MN}$ being the generalized energy-momentum tensor of the
scalar field defined as
\bea
\mathbb{T}^{(\Phi)}_{MN}= \nabla_{M}\Phi\nabla_{N}\Phi - g_{MN}\bigl[\frac{1}{2}(\nabla\Phi)^2
+V(\Phi)\bigr] + \frac{1}{\kappa^2_5}\left[\nabla_{M}\nabla_{N}f(\Phi) - g_{MN}\,\nabla^2f(\Phi)\right]. \label{generalized}
\eea
{

In order to derive the explicit form of the above field equations, we need to
combine the non-vanishing components of the energy-momentum tensor with those
of the Einstein tensor $G_{MN}$ presented in Eqs. (\ref{Ein-diag-comp})-(\ref{Ein-yy-comp}). First, the
off-diagonal components $\mathbb{T}^{y}_{\ r}$, $\mathbb{T}^{v}_{\ r}$,
$\mathbb{T}^{y}_{\ v}$, $\mathbb{T}^{r}_{\ v}$ lead, respectively,
to the following four equations:
\bea
(1+f'')\,\partial_{y}\Phi\,\partial_{r}\Phi +
f'\,\partial_{y}\partial_{r}\Phi-
A' f'\,\partial_{r}\Phi =0 \,, \label{conf-yr-comp}
\eea
\vskip -0.8cm
\bea
(1+f'')\,(\partial_{r}\Phi)^2 +
f'\,\partial^{2}_{r}\Phi =0 \,, \label{conf-vr-comp}
\eea
\vskip -0.8cm
\bea
(1+f'')\,\partial_{y}\Phi\,\partial_{v}\Phi + f'\,\partial_{y}\partial_{v}\Phi - 
A' f'\,\partial_{v}\Phi - \frac{\partial_{y}m}{r}\,f'\,\partial_{r}\Phi 
= f\,\frac{\partial_{y}m}{r^2}\,, \label{conf-yv-comp}
\eea
\vskip -0.2cm
\bea
&& \hspace*{-0.5cm}(1+f'')\,(\partial_{v}\Phi)^2 + f'\,\partial^{2}_{v}\Phi - 
\frac{m}{r^2}\,f'\,\partial_{v}\Phi-\frac{\partial_{v}m}{r}\,f'\,\partial_{r}\Phi
+ e^{2A}\,\frac{\partial_{y}m}{r}f'\,\partial_{y}\Phi +\nonumber\\[0.5mm]
&& \hspace*{-0.5cm} (1-\frac{2m}{r})\bigl[(1+f'')\,\partial_{v}\Phi\,\partial_{r}\Phi +
f'\,\partial_{v}\partial_{r}\Phi\bigr]
= f\Bigl[\frac{2}{r^2}\,\partial_{v}m - 
\frac{e^{2A}}{r}\,(\partial^{2}_{y}m + 4A'\partial_{y}m)\Bigr]. \label{conf-rv-comp}
\eea
In the above, $f'$ and $f''$ denote the first and second, respectively, derivative
of the coupling function $f$ with respect to $\Phi$, and, for simplicity, ${\kappa^2_{5}}$
has been set to unity. Also, note that the off-diagonal components of the energy-momentum
tensor $\mathbb{T}^{v}_{\ y}$ and $\mathbb{T}^{r}_{\ y}$ are
not independent and their corresponding equations reduce again to Eqs. (\ref{conf-yr-comp})
and (\ref{conf-yv-comp}).

Furthermore, the diagonal components provide us with three additional equations:
\bea
&& \hspace*{-2cm} e^{-2A}\bigl[(1+f'')\,\partial_{v}\Phi\,\partial_{r}\Phi + 
f'\,\partial_{v}\partial_{r}\Phi + \frac{m}{r^2}\,f'\,\partial_{r}\Phi\bigr]
+ A'f'\,\partial_{y}\Phi \nonumber\\[1mm] && \hspace*{4cm}
-(\mathcal{L}_{\Phi} + \Box f + \Lambda_{B})= 3 f\,(2A'^2 + A'') \,, \label{vv-comp}
\eea
\bea
\frac{e^{-2A}}{r}\,f'\Bigl[\partial_{v}\Phi +\bigl(1-\frac{2m}{r}\bigr)\partial_{r}\Phi\Bigr]
+ A'f'\,\partial_{y}\Phi-(\mathcal{L}_{\Phi} + \Box f + \Lambda_{B})= 3 f\,(2A'^2 + A'')\,,
\label{thth-comp}
\eea
\bea
(1+f'')\,(\partial_{y}\Phi)^2 + f'\,\partial^2_{y}\Phi
 -(\mathcal{L}_{\Phi} + \Box f + \Lambda_{B}) = 6f\,A'^{2}\,. \label{yy-comp}
\eea
The above equations contain the complicated expressions of $\mathcal{L}_{\Phi}$
and $\Box f$, which are given by
\beq
\mathcal{L}_{\Phi} \equiv \frac{1}{2}(\nabla\Phi)^2+V(\Phi)=\frac{e^{-2A}}{2}\Bigl[
2\,\partial_v\Phi\,\partial_r\Phi+\bigl(1-\frac{2m}{r}\bigr)(\partial_r\Phi)^2\Bigr]
+\frac{1}{2}\,(\partial_y\Phi)^2 +V(\Phi)\,,
\eeq
and
\beq
\Box f 
= e^{-2A}\,\partial_v\partial_rf+\frac{e^{-2A}}{r^2}\,\partial_r\Bigl[
r^2 \partial_vf+r^2\bigl(1-\frac{2m}{r}\bigr)\,\partial_rf\Bigr]
+e^{-4A}\partial_y\bigl(e^{4A}\,\partial_yf\bigr)\,,
\eeq
respectively, and are thus cumbersome to use. However, the combination of
Eqs. (\ref{vv-comp}) and (\ref{thth-comp}) results in a simpler and more useful
condition, namely
\bea
(1+f'')\,\partial_{v}\Phi\,\partial_{r}\Phi + f'\,\partial_{v}\partial_{r}\Phi=
\frac{f'}{r}\,\Bigl[\partial_{v}\Phi + (1-\frac{3m}{r})\,\partial_{r}\Phi\Bigl]\,.
\label{constraint}
\eea

In the above analysis, we have once again assumed that the scalar field $\Phi$, and
consequently the coupling function $f$, does not depend on the angular coordinates $\theta$ and
$\phi$ in order to preserve the spherical symmetry of the solutions on the brane.
We have nevertheless retained their dependence on all remaining coordinates
($r,v,y$). It is easy to see that any simpler ansatz fails to pass the field
equations: if we assume that the scalar field $\Phi$ depends only on the 
bulk coordinate $y$, then Eq.  (\ref{conf-yv-comp}) leads to the result
$\partial_ym=0$ -- the same equation is inconsistent due to its explicit
$r$-dependence in the case where $\Phi$ is assumed to be only a function
of the time-coordinate $v$; finally, if the field depends only on the radial
coordinate $r$, then Eq. (\ref{conf-yr-comp}) demands that $f'=0$ - but
this takes us back to the minimal-coupling case that has already been 
excluded \cite{KOT}. 

The above arguments clearly indicate that the scalar field $\Phi$ must
depend at least on a pair of coordinates. In fact, even the assumption that 
it depends on only two coordinates is inconsistent with the field equations, since:
\begin{itemize}
\item{} if $\Phi=\Phi(v,y)$ and thus $\partial_r\Phi=0$, Eq. (\ref{constraint}) 
leads to either $\partial_v\Phi=0$ (excluded above) or $f'=0$ -- but the
latter option again reduces Eq. (\ref{conf-yv-comp}) to an inconsistent equation.
\item{} if $\Phi=\Phi(v,r)$ and thus $\partial_y\Phi=0$, Eq.~(\ref{conf-yr-comp})
demands, for $\partial_r\Phi \neq0$, $f'=0$ -- then, Eq. (\ref{constraint})
leads to $\partial_v\Phi=0$ which is in contradiction with our assumption.
\item{} if $\Phi=\Phi(r,y)$ and thus $\partial_v\Phi=0$, Eq. (\ref{constraint})
demands, for $\partial_r\Phi \neq0$, $f'=0$ -- then, Eq.~(\ref{conf-yr-comp})
leads to $\partial_y\Phi=0$ which is again in contradiction with our assumption.
\end{itemize}
Therefore, we conclude that any attempted simplification in the form of the scalar
field does not conform with the field equations, and this, interestingly enough,
holds regardless of the form of the coupling function $f(\Phi)$.
We are thus led to consider whether the only remaining possibility $\Phi(r,v,y)$,
in conjunction with an appropriate choice of $f(\Phi)$, could support the existence
of a solution with a mass function $m = m(v,y)$ that would perhaps localise a
black hole together with its singularity close to the brane. Therefore, in
what follows we consider a number of natural choices for the coupling function
$f(\Phi)$ and investigate whether these can lead to any viable solutions.

\subsection{The $f(\Phi)= a\,\Phi$ case}

Postulating that $f(\Phi)=a\,\Phi$, with $a$ being a constant, gives $f'(\Phi)=a$
and $f''(\Phi)=0$, which significantly simplifies the field equations. Looking for 
a solution for $\Phi(v,r,y)$, we immediately see that a purely factorised form,
e.g. $\Phi(v,r,y)=U(v) R(r) Y(y)$, or any other form in which at least one of
the coordinates is factorised out, are excluded as they fail to satisfy the field
equations. 

As a matter of fact, for the particular choice of the coupling function $f$,
Eq. (\ref{conf-vr-comp}) can be analytically integrated to determine the form of
$\Phi$. For $f(\Phi)=a\,\Phi$, it takes the form
\beq
\frac{\partial_r^2\Phi}{(\partial_r \Phi)^2}=-\frac{1}{a}\,,
\eeq
and, upon integrating twice, it yields the general solution
\beq
\Phi(v,r,y)= a\,\ln\,[r +a B(v,y)]+C(v,y)\,,
\eeq
where $B(v,y)$ and $C(v,y)$ are arbitrary functions. However, the above solution
fails again to satisfy the condition (\ref{constraint}): this takes the form
$a\,\partial_v B+B\,\partial_v C+1-3m/r=0$ that cannot be satisfied due to the
explicit dependence on $r$. This result therefore excludes
the particular choice for the coupling function.


\subsection{The $f(\Phi)=a\,\Phi^2$ case}

Also in this case, upon substituting $f'(\Phi)=2a\Phi$ and $f''(\Phi)=2a$, where
$a$ is again a constant, Eq. (\ref{conf-vr-comp}) takes the form
\beq
-\frac{(1+2a)}{2a}\,\frac{\partial_r \Phi}{\Phi}=\frac{\partial_r^2 \Phi}
{\partial_r \Phi}\,.
\eeq
This can be analytically integrated twice to yield the general solution for $\Phi$,
namely
\beq
\Phi(v,r,y)=[B(v,y)\,r +C(v,y)]^{2a/(1+4a)}\,,
\eeq
where again $B(v,y)$ and $C(v,y)$ are arbitrary functions. Interestingly enough, the
above form of the scalar field together with the assumption $f(\Phi)=a\,\Phi^2$ 
manage to satisfy all off-diagonal equations (\ref{conf-yr-comp})-(\ref{conf-rv-comp}), with
the latter providing constraints that determine the unknown functions $B(v,y)$ and
$C(v,y)$ in terms of the warp factor $A(y)$ and the mass function $m(v,y)$. 
However, the diagonal equations (\ref{vv-comp})-(\ref{yy-comp}) are more
difficult to satisfy with the constraint (\ref{constraint}) proving the
particular configuration of $f$ and $\Phi$ once again inconsistent by taking
the form $\partial_v C +B (1-3m/r)=0$ and thus demanding
the trivial result $B(v,y)=0$. 


\subsection{The $f(\Phi)=a\,\Phi^n$ case}

In this case, we have $f'(\Phi)=an\,\Phi^{n-1}$ and $f^{''}(\Phi)=an(n-1)\,\Phi^{n-2}$,
and Eq. (\ref{conf-vr-comp}) takes the form
\beq
-\frac{1}{an}\,[\Phi^{1-n}+\frac{an (n-1)}{\Phi}]\,\partial_r \Phi=
\frac{\partial_r^2 \Phi}{\partial_r \Phi}\,.
\eeq
Integrating the above, we obtain
\beq
\partial_r\Phi(v,r,y)=b(v,y)\,\Phi^{1-n}\,\exp\biggl[\frac{\Phi^{2-n}}{an(n-2)}\biggr]\,,
\label{diffPhi-n}
\eeq
where $b(v,y)$ an arbitrary function. Unfortunately, the solution of the above 
first-order differential equation for $n \geq 3$ cannot be written in a closed form.
However, the following integral form
\beq
\int\,d\Phi\,\Phi^{n-1}\,\exp\biggl[-\frac{\Phi^{2-n}}{an(n-2)}\biggr]=
b(v,y)\,r+c(v,y)\,,
\eeq
where $c(v,y)$ is another arbitrary function, will prove to be more than adequate
for our purpose. Although an explicit form for the scalar field $\Phi$ cannot be found,
differentiating both sides of the above equation with respect to $v$ yields
\beq
\partial_v\Phi(v,r,y)=\Phi^{1-n}\,\exp\biggl[\frac{\Phi^{2-n}}{an(n-2)}\biggr]\,
\left[\partial_v b(v,y)\,r+\partial_v c(v,y)\right].
\eeq
Differentiating also Eq. (\ref{diffPhi-n}) with respect to $v$ yields 
$\partial_v\partial_r \Phi$ and upon substitution of the relevant quantities
in Eq. (\ref{constraint}), we obtain once again the, condemning for our ansatz,
constraint $\partial_v c +b (1-3m/r)=0$. 

It is worth noting that the case where the coupling function $f(\Phi)$ is a 
linear combination of different powers of $\Phi$, i.e. $f(\Phi)=\sum_{k=0}^n
a_k\,\Phi^k$, was also considered \footnote{This particular choice for the
coupling of a bulk scalar field to the Ricci scalar was considered in \cite{BDT}
in the context of a brane-world cosmological solution that could produce
accelerated expansion on the brane at late times.}. For $n=1$ and $n=2$, the analyses followed
closely the ones for the cases with $f(\Phi)=a \Phi$ and $f(\Phi)=a \Phi^2$,
respectively, leaving no space for a viable solution. For $n=3$, Eq. (\ref{conf-vr-comp})
could be again integrated once to yield the result
\beq
\partial_r\Phi(v,r,y)=\frac{b(v,y)}{a_1+2a_2\Phi+3a_3\Phi^2}\,
\exp\biggl[-\frac{1}{\lambda}\,\arctan\Bigl(\frac{a_2+3a_3\Phi}{\lambda}\Bigl)\biggr]\,,
\eeq
where $\lambda=\sqrt{3a_1a_3-a_2^2}$. Integrating once more, we obtain again an
integral equation. Following a similar analysis as above, we arrive again, from
Eq. (\ref{constraint}), at the constraint $\partial_v c +b (1-3m/r)=0$ and the
trivial result $b(v,y)=0$. For $n \geq 4$, our set of equations do not give
a closed form even for $\partial_r \Phi$.


\subsection{The $f(\Phi)=e^{k \Phi}$ case}

We finally consider the case of an exponential coupling function for which
$f'(\Phi)=k\,e^{k \Phi}$ and $f''(\Phi)=k^2\,e^{k \Phi}$, where $k$
is a constant -- note than an arbitrary constant multiplying the
exponential function can be absorbed into the value of $\Phi$ and thus is
set to unity. Then, Eq. (\ref{conf-vr-comp}) takes the form
\beq
-\frac{1}{k}\,\bigl(e^{-k \Phi}+k^2\bigr)\,\partial_r \Phi=
\frac{\partial_r^2 \Phi}{\partial_r \Phi}\,,
\eeq
with solution
\beq
\partial_r\Phi(v,r,y)=b(v,y)\,e^{-k \Phi}\,\exp\biggl[\frac{e^{-k \Phi}}{k^2}\biggr].
\label{diffPhi-exp}
\eeq
Integrating once more, we obtain
\beq
\int\,d\Phi\,e^{k \Phi}\,\exp\biggl[-\frac{e^{-k \Phi}}{k^2}\biggr]=
b(v,y)\,r+c(v,y)\,. \label{integral-exp}
\eeq
Deriving, from Eqs. (\ref{diffPhi-exp}) and (\ref{integral-exp}), the expressions for
$\partial_v\partial_r \Phi$ and $\partial_v\Phi$, respectively, and substituting them
together with $\partial_r \Phi$ in Eq. (\ref{constraint}), we obtain again
the constraint $\partial_v c +b (1-3m/r)=0$, that clearly excludes the exponential
ansatz as well. 


\subsection{A general no-go argument}

The failure of finding a viable solution, after a variety of forms for the
coupling function $f(\Phi)$ have been considered, seems to hint that perhaps
a theory of a non-minimally-coupled scalar field is altogether inconsistent with
the realisation of the additional bulk matter  necessary to support a spacetime
described by the line element~(\ref{metric}). 
In that case, one should be able to develop a general argument
that would exclude the emergence of a solution independently of the form of
the coupling function $f(\Phi)$. 

To this end, we bring Eq. (\ref{conf-vr-comp}) to the form
\begin{equation} 
1+f''(\Phi)=-f'(\Phi)\,\frac{\partial_r^2\Phi}{(\partial_r\Phi)^2}\,,
\label{f''}
\end{equation}
which we can replace into Eq. (\ref{conf-yr-comp}) to obtain
\begin{equation}
A'=\partial_r\left(\frac{\partial_y\Phi}{\partial_r\Phi}\right)\,.
\end{equation}
The above differential equation can be integrated with respect to $r$ to give
\begin{equation}
\partial_y\Phi=\partial_r\Phi \bigl[A'(y)\,r+F(v,y)\bigr]\,.
\label{sol_y}
\end{equation}
Similarly, Eq. (\ref{constraint}) can be brought to the following form
\begin{equation}
\left(\partial_r-\frac{1}{r}\right)\frac{\partial_v\Phi}{\partial_r\Phi}=
\frac1r\left(1-\frac{3m}{r}\right),
\end{equation}
which upon integration with respect to $r$ yields
\begin{equation}
\partial_v\Phi=\partial_r\Phi\,\Bigl[-1+\frac{3m}{2r}+D(v,y)\,r\Bigr]\,.
\label{sol_v}
\end{equation}
The functions $F(v,y)$ and $D(v,y)$ appearing in Eqs. (\ref{sol_y}) and
(\ref{sol_v}) are, at the moment, completely arbitrary. It can, however, 
be easily checked that there exists a relation between them. To establish
 this relation, we proceed as follows. First, we differentiate Eq.~(\ref{sol_y}) 
with respect to $v$ and Eq. (\ref{sol_v}) with respect to $y$ to obtain
\begin{eqnarray}
\partial_v\partial_y\Phi&=&\partial_vF(v,y)\,\partial_r\Phi+
\left(A'r+F\right)\,\partial_r\partial_v\Phi\,, \label{double_diff_1} \\
\partial_y\partial_v\Phi&=&r\partial_y D(v,y)\,\partial_r\Phi+ 
\left(-1+\frac{3m}{2r}+D(v,y)\,r\right)\partial_r\partial_y\Phi\,.
\label{double_diff_2}
\end{eqnarray}
Equating the right-hand sides of the above two equations, we arrive at
the relation
\beq
\partial_vF(v,y)\,\partial_r\Phi+(A'r+F)\,\partial_r\partial_v\Phi=
r\partial_y D(v,y)\,\partial_r\Phi+ \left(-1+\frac{3m}{2r}+D(v,y)\,r\right)
\partial_r\partial_y\Phi\,. 
\eeq
Taking finally the derivatives of Eqs. (\ref{sol_y}) and (\ref{sol_v}) with
respect to $r$, these yield the expressions of the double derivatives
$\partial_r\partial_y\Phi$ and $\partial_r\partial_v\Phi$ that appear above.
Substituting and simplifying leads to the final constraint
\beq
-\frac{3m}{2r^2}\,F(v,y)-\frac{3m}{r}\left(A'+\frac{\partial_ym}{2m}\right)+
\partial_vF(v,y)+A'+F(v,y)D(v,y)-r\partial_yD(v,y)=0\,.
\label{final}
\eeq
However, the above is catastrophic for the existence of the desired solution.
The only way the above relation can hold is if the coefficients of all powers
of $r$ identically vanish. This leads to the result that $F(v,y)=0$, that
subsequently demands that $A'(y)=0$ which is clearly in contradiction with
our assumption as it eliminates the warp factor from the model. In addition,
the desired dependence of the mass term on the extra coordinate $y$ is also
forced to vanish, once we assume that $A'(y)=0$, which destroys the localisation
of the black-hole singularity.
 
Although of a secondary importance, let us finally note that even if 
the function $A(y)$ were not forced to be trivial, the constraint following
from the second term of Eq.~(\ref{final}) would lead to the result 
$m(v,y) \sim e^{-2A(y)}$ -- thus, for a decreasing warp factor,
the mass term would have to increase away from the brane thus
invalidating the idea of the localisation of black hole. Therefore,
a viable field-theory model should not only support a non-trivial
profile of the mass function of the black hole but also a profile that
could localise the black hole close to the brane.     


\section{Discussion and Conclusions}

Despite an intensive research activity over a period of almost fifteen years, a closed-form 
analytical solution that would describe a 5-dimensional regular black hole localised on a brane is still
missing. Although numerical solutions that reassure us of their existence have appeared in the literature,
 the way to proceed in order to derive a complete analytical solution remains unclear.
As almost all of those numerical solutions rely on the presence of
some type of matter, either on the brane or in the bulk, in this work, we
turned to a previous idea, introduced by one of the authors and collaborators, that a type
of bulk matter can help to localise the extended black-string singularity close to the brane
and thus restore the finiteness of the 5-dimensional AdS spacetime at a small distance
from the brane. 

However, the metric ansatz that would describe a 5-dimensional spacetime of this form had
to be carefully constructed. The black-string spacetime was associated to a factorised metric
ansatz, therefore, the localisation of the extended singularity would be realised only through
a non-factorised ansatz, in which the 4-dimensional part would exhibit dependence on the
fifth coordinate. Previous attempts \cite{KT, KOT} had shown that such line-elements characterised
by the presence of a horizon in their 4-dimensional part led to spacetimes with additional
singularities apart from the extended black-string one. A modified Vaidya-type 4-dimensional
line-element was finally chosen and embedded in a 5-dimensional warped spacetime. 
Being analytic in 4 dimensions, this metric ansatz was free from any additional singularities.
Moreover, its mass being a function of both the fifth and the time-coordinate,
it provided a reasonable ansatz for a perturbed Schwarzschild background on the brane,
ideal for investigating both the localisation of the black-hole singularity and the existence
of a static solution.   

The gravitational part of our model decided, we  turned to the determination of
the field theory model that would support such a spacetime. Previous attempts to find
such a model based on ordinary theories of scalar or gauge fields had led to a negative
result \cite{KT}. Therefore, in this work, we decided to study instead a variety of field
theories that could be characterised as non-ordinary -- for simplicity, we focused on
the case of scalar field theories. In Section 3, we examined the case of a field theory
with one or more scalar fields minimally-coupled to gravity but otherwise described by
a general Lagrangian. The cases studied included a single scalar field with a non-canonical
kinetic term and two interacting scalar fields with either canonical, non-canonical or
mixed kinetic terms. Our analysis allowed for general forms of potentials as well as the
case where one or both of the scalar fields were ghosts. In Section 4, we turned to the 
field theory of a single scalar field non-minimally-coupled to gravity, and studied
the cases where its coupling function was a power-law of the field, a polynomial, 
an exponential function, or of a completely arbitrary form. 

In order to avoid any unreasonable restrictions on the field configurations, we allowed
the warp factor to assume a $y$-dependent, but otherwise arbitrary form. We also
imposed no fine-tunings between bulk and brane parameters. A viable bulk
solution, if emerged, would be subsequently used, to determine, through the
junction conditions, the brane content. Nevertheless, our analysis never reached that
point: all the field theory models studied, no matter how general, were shown not to
be able to support the assumed gravitational background. Considering only the set of
gravitational equations in the bulk, we were able to demonstrate that in each and every
case, the scalar field-theory model chosen was not compatible with the basic assumptions
for the metric ansatz necessary for the localisation of the black-string singularity.

Our analysis has, nevertheless, confirmed that such a localisation demands the synergetic 
action of both the bulk and the brane part of spacetime. The chosen metric ansatz
introduces in the bulk, apart from an energy-density and an isotropic diagonal pressure that
satisfy a stiff equation of state, additional off-diagonal, non-isotropic pressure components.
The dependence of the mass function on both the fifth- and the time-coordinate contributes
to these. It becomes therefore clear that gravitational degrees of freedom tend to leak from the
brane -- similarly to the black-hole singularity -- and, although the models considered
in this work have failed to localise them, a mechanism must exist that will achieve this.
Another important point that has emerged from our analysis is the necessity of
the time-dependence of the field configurations in all the models we studied -- even when
the mass parameter is assumed to be time-independent; according to our findings, a
static black-hole configuration may indeed exist, however, the associated field configuration
itself has to be dynamic. 

In the previous related work \cite{KT}, configurations involving also gauge fields were studied;
the arguments however that excluded the existence of a viable solution were identical to
the ones used for the case of scalar models. Although, in the present work, we have restricted
our study in scalar field-theory models, we anticipate that similar results would follow
even in the case of non-ordinary gauge field-theory models -- we have postponed this
study for a future work. Finally, one should note that all of the above observations are
independent of the sign of the parameter $\epsilon$ that appears in our metric
ansatz, and thus hold not only for the creation of a brane-world black hole but also
for any expanding distribution of matter in a brane-world set-up. 

Our analysis is by no means exhaustive. Nevertheless, in our attempt to generate the
bulk energy-momentum tensor necessary for the localisation of the black-hole topology
close to the brane, we have considered a general selection of non-ordinary scalar
field-theory models with a high degree of flexibility, and reached a negative result in
each case. We have also considered a particular non-factorised metric ansatz -- 
no matter how well motivated this choice was, we cannot exclude the possibility that
the 5-dimensional line-element assumes a different form that may perhaps be related
to the Schwarzschild black-hole metric on the brane in a more subtle way
(see, for example, the construction of brane-localised black holes in the
lower-dimensional case \cite{EHM, AS} based on the use of a C-metric -- there, a
Schwarzschild-like metric for the geometry on the brane was derived, however
it was not a vacuum solution). Our
results demonstrate how difficult, if at all possible, the construction of a localised 
5-dimensional black hole may be in the context of a well-defined field-theory model.


{\bf Acknowledgments} P.K and K.Z.E would like to thank Kyriakos Tamvakis for illuminating
discussions during the early stages of this work. This research has been co-financed by the
European Union (European Social Fund - ESF) and Greek national funds through the Operational
Program ``Education and Lifelong Learning'' of the National Strategic Reference Framework
(NSRF) - Research Funding Program: ``ARISTEIA. Investing in the society of knowledge through
the European Social Fund''. Part of this work was supported by the COST Actions MP0905
``Black Holes in a Violent Universe'' and MP1210 ``The String Theory Universe''.

\end{document}